\documentclass[aps,prd,reprint,amsmath,amsfonts,nofootinbib,longbibliography]{revtex4-1}

\usepackage[T1]{fontenc}
\usepackage[utf8]{inputenc}
\usepackage{braket}
\usepackage{graphicx}
\usepackage{siunitx}

\usepackage{xcolor}

\newcommand{\diff}{\mathrm{d}}
\newcommand{\e}{\text{e}}
\newcommand{\im}{\text{i}}
\newcommand{\prik}{\boldsymbol{\cdot}}
\newcommand{\mathand}{\quad\text{and}\quad}

\begin{document}

\title{The Real-Time Correlation Function of Floquet Conformal Fields}

\author{Malthe Andersen}
\affiliation{Institut for Fysik og Astronomi, Aarhus Universitet, Denmark}

\author{Frederik N{\o}rfjand}
\affiliation{Institut for Fysik og Astronomi, Aarhus Universitet, Denmark}

\author{Nikolaj Thomas Zinner}
\affiliation{Institut for Fysik og Astronomi, Aarhus Universitet, Denmark}

\date{\today}

\begin{abstract}
Conformal field theories (CFT) play an important role in quantum field theories as they allow for exact studies that are hard to access without the conformal symmetries in place. A remarkable recent development is the application of conformal field theory to periodically driven systems, the so-called Floquet-driven CFTs. A number of recent works have found that there are classes of periodically driven (1+1)-dimensional CFTs that can be analytically solved, and that show both heating and non-heating phases, as well as unusual properties in terms of information scrambling. Hitherto, most results have focused on the dynamics of such models at stroboscopic times, i.e. samples taken at each period of the time evolution. In this paper, we consider time-evolution at arbitrary real times and derive formulas for the exact time-propagation of heating and non-heating classes of Floquet-driven CFTs. This requires a subtle treatment of non-stroboscopic time that leads to different interpretations in terms of curved space-time dynamics in these systems compared to previous results in the literature.
\end{abstract}
\maketitle

\twocolumngrid

	\subsubsection*{Introduction}
\addcontentsline{toc}{section}{\protect Introduction}
Floquet driving is a popular subject of investigation. It provides a setting that is useful for studying the principles of statistical dynamics beyond equilibrium, and it provides an arena for studying exotic systems with behavior that cannot be realized in non-driven systems. These include Floquet topological phases \cite{kitagawa2010topological, jiang2011majorana, rudner2013anomalous,von2016phase, else2016classification, potter2016classification, roy2016abelian, po2016chiral, roy2017periodic, harper2017floquet, po2017radical, potirniche2017floquet, morimoto2017floquet, fidkowski2019interacting, yang2020photonic, wu2020floquet, hu2020dynamical,  rubio2020floquet} and time crystals \cite{khemani2016phase, else2016floquet, von2016phase2, else2017prethermal, yao2017discrete, kozin2019quantum, surace2019floquet, yao2020classical}. A recent review of the developments in Floquet topological phases is found in \cite{harper2020topology}. The interest in Floquet driven systems goes hand in hand with advances in experimental techniques which allow the study of Floquet systems in the laboratory \cite{reitter2017interaction, zhang2017observation, choi2017observation, maczewsky2017observation, tang2018thermalization, li2019observation, mukherjee2020observation}.

A Floquet driven system is defined by having a time-periodic Hamiltonian $H(t)=H(t+T)$ for some fixed period $T$. Systems, where this symmetry is spontaneously broken, are called discrete time crystals. This is realized when a system's ground state does not exhibit the time-periodicity of its Hamiltonian.

This paper concerns itself with Floquet driven conformal field theories. Conformal field theories (CFT's) are symmetric under the very restrictive set of conformal transformations. Due to this set of symmetries, conformal fields are extraordinarily simple compared to usual quantum field theories. The references \cite{LectCFT, PaulGinspargCFT, IntroCFTBook} introduces the reader to conformal transformations and conformal field theories.

In 2018, Wen and Wu found that (1+1)-dimensional conformal field theories play an important role in the study of Floquet driven systems \cite{wen2018floquet, wen2018quantum}. They constructed a model which has an exactly solvable Floquet driving protocol. This is rare since numerical methods are usually needed.

The Floquet driving protocol that Wen and Wu developed in conformal field theory utilizes a periodic change between doing time-evolution with a free conformal Hamiltonian and the sine-square deformed (SSD) conformal Hamiltonian. Sine-square deformation was initially developed in relation to $N$-site one-dimensional quantum systems as a numerical trick. When deriving thermodynamic properties of such systems, the effects of boundary conditions are usually not wanted. Hence such systems are mostly imposed periodic boundary conditions. Though, it so happens that some numerical methods work best with open boundary conditions. Hence it is preferable to have a boundary condition suppression mechanism. It was found numerically that certain open quantum systems behave as their periodic counterpart when subject to a sine-square deformation of their coupling strengths. 

Sine-square deformation (SSD) was first introduced by Gendiar \textit{et al.} in 2009 \cite{gendiar2009spherical} in relation to a one-dimensional spinless fermion chain. Hikihara \textit{et al.} \cite{hikihara2011connecting} considered SSD in relation to spin chains, and found numerically the groundstate wave-functions to agree almost exactly between the SSD open boundary and the periodic boundary. Other models like the Hubbard model and the Kondo-lattice model has been investigated using SSD  \cite{gendiar2011suppression,shibata2011boundary}. In 2011 Katsura \cite{katsura2011exact} found analytically that the groundstate energies agree exactly between a SSD fermion chain and the same chain with periodic boundaries. This is quite a remarkable result considering the different topologies of the systems. It shows that SSD is a bridge between topologies, giving an effective circular topology to an open system and vice versa. It was argued in \cite{maruyama2011sine}, the same year, that the equal groundstate energy result holds in higher dimensions as well.  

The conformal SSD theory with implications to our work was published in 2012 by Katsura \cite{katsura}. He showed for the first time that SSD fits naturally into a conformal field theory by rewriting the conformal SSD Hamiltonian in terms of Virasoro operators. It was found that the vacuum of a periodic conformal field theory is an eigenstate of the same energy in the SSD theory due to translational symmetry. The same is not true for the open boundary groundstate. We make note that a sine-square deformed theory is insensitive to the boundary conditions of a theory. It suppresses the Hamiltonian density at the endpoints, and thus both open and closed boundaries give the same SSD theory. 

Further investigations considering SSD conformal field theories and their quantization were done in \cite{ishibashi2015infinite, ishibashi2016dipolar} by Ishibashi and Tada who found a polar quantization method. They found the conformal SSD Hamiltonian to have a continuous spectrum and an infinite length scale, though being defined on a finite space. Later the Möbius quantization was developed in \cite{okunishi2016sine}. Further work has been done in \cite{wen2016evolution, tamura2017zero}. In \cite{tada2015sine}, SSD is introduced into closed string theory via a diverging worldsheet metric. Recently (2020), Lapierre and Moosavi \cite{GeneralDeformations} proposed a method for incorporating more general deformations into conformal Floquet dynamics, making the SSD Hamiltonian into a special case. This new approach allows a variety of new possibilities in the field.

The Conformal Floquet theory developed by Wu and Wen has both a heating and a non-heating phase characterized by a growing or periodic von Neumann entropy, respectively. Consider \cite{fan2020emergent} for a discussion of energy and entanglement considerations in these phases. At the present date (2020), work is done on generalizing Wu and Wen's model \cite{GeneralDeformations, fan2020floquet, han2020classification}. 

Lapierre \textit{et al.} \cite{lapierre2020emergent} (2020) found the astonishing result that excitations, defined by the Floquet driven conformal two-point functions, move as quasi-particles in a (1+1)-dimensional spacetime with two Schwarzschild black holes. This result is based on the assumption that the equations obtained from $T$-periodic driving can be extrapolated to all real times. 
However, here we present a different method to calculate the two-point function that applies for all real times and therefore encaptures the micro-motion of the system. We find different
results as compared to \cite{lapierre2020emergent} when we are away from times that can be written as an integer times
the period of the Hamiltonian, i.e. at times that are not stroboscopic. We identify the difference to the 
treatment of the free Hamiltonian part of the time-development, and using the exact method, we only recover behavior 
akin to a black hole geometry for stroboscopic times. For all other times, we find markedly different dynamics. 

We discuss the delicacy of extrapolating to real times, and show the results of the exact method, considering both the heating and the non-heating phase. We are thus forced to conclude that, while the excitations still move as quasi-particles in a spacetime that 
is effectively curved, the question of whether this allows the study of particular black hole setups in laboratories using 
Floquet conformal dynamics remains elusive. The resulting spacetime is of a more complicated nature and we have not found a 
simple analytical treatment to extract its exact features. This remains a very interesting research question for the immediate 
future.

\subsubsection*{Conformal Floquet dynamics}
\addcontentsline{toc}{section}{\protect Conformal Floquet dynamics}
For a 1-dimensional\footnote{Often called (1+1)-dimensional since there is one spacial and one temporal dimension.} CFT on the interval $[0,l]$ consider the Hamiltonian density $\mathcal{H}$ of the field as well as the ordinary Hamiltonian $H_0$:
	\begin{align}
		H_0=\int_{0}^{l}\mathcal{H}\diff x
	\end{align}
	The sine square deformation (SSD) of the field is then given by (discussed in more detail in \cite{katsura}):
	\begin{align}
		H_{\text{SSD}}=\int_{0}^{l}2\sin^2\left(\frac{\pi x}{l}\right)\mathcal{H}\diff x
	\end{align}
	It is possible to drive the system by changing between the ordinary Hamiltonian and the sine square deformed Hamiltonian periodically by extending the following procedure to all times:
	\begin{align}
		H=\left\{\begin{matrix}
		H_0&\text{for}&0\le t< T_0\\H_{\text{SSD}}&\text{for}&T_0\le t< T_0+T_1
		\end{matrix}\right.
		\label{Hamiltonian}
	\end{align}
	That is when the first driving period $T=T_0+T_1$ has passed the system is again described by $H_0$ during the time span $T_0$ followed by $H_{\text{SSD}}$ during the span $T_1$ and so on. The shift from one Hamiltonian to the other is assumed to be instantaneous which in practice means that the shift should be short compared to the general order of time in the system. That is, compared to the periods $T_0$ and $T_1$, such that the transition periods have negligible effect. This type of driving has been investigated in several articles such as \cite{wen2018floquet,wen2018quantum,lapierre2020emergent,EmergentSpacial}. A generalization of the drive to quasi-periodic driving can be seen in \cite{quasiperiodic}. The goal is to find the two-point correlation function\footnote{The theory can also be extended to $n$-point functions, but this will not be discussed here.} of some primary field $\phi$ with conformal weights $h$ and $\overline{h}$ denoted $G(x,t,x_0,0)$ for all $t\in\mathbb{R}$. Sometimes, it is necessary to understand the complete real-time dynamics of the system and not just the results at stroboscopic times $t=nT$ for some non-negative integer $n$. For instance, this would be important if one were to investigate the effects of uncertainties on the driving periods. Experimentally, the system would be driven not by $T_0$ and $T_1$ exactly, but by small deviations from these distributing around $T_0$ and $T_1$, and this could have important consequences. It is also important for the interesting relation to black holes found in \cite{lapierre2020emergent} as will be discussed briefly later.
	
	Now we a short overview of how to obtain the two-point function at stroboscopic times. The first step is to find the two-point function after one driving cycle:
	\footnotesize
	\begin{align}
		G(x,T,x_0,0)=\left<\e^{\im T_1 H_{\text{SSD}}}\e^{\im T_0 H_0}\phi(x,0)\e^{-\im T_0 H_0}\e^{-\im T_1 H_{\text{SSD}}}\phi(x_0,0)\right>
	\end{align}
	\normalsize
	 Now the coordinates $x$ and $t$ are changed to be more convenient. First, define the imaginary time $\tau=\im t$ and then change the coordinates to:
	\begin{align}
		\omega=\tau+\im x\mathand \overline{\omega}=\tau-\im x
	\end{align}
	It might appear contradictory, but assume for now that $\tau$ is real-valued. Whenever a conclusion is to be drawn, the actual imaginary value of $\tau$ should then be inserted into the equations. Thus, the change to $\omega$-coordinates is a change to a strip of height $l$ in the complex plane. Next, change coordinates to the entire complex plane using the exponential function\footnote{Only the expressions for the first coordinate is given from now on, but the formulas for the other coordinate are easily guessed from those. Also notice that $\overline{\enskip\prik\enskip}$ does not denote the complex conjugate in this text.}:
	\begin{align}
		z=\e^{\frac{2\pi}{l}\omega}
	\end{align}
		Let $z_0$ and $\overline{z}_0$ denote the exponential coordinate transformation from the coordinates $x=x_0$ and $t=0$. It was found in \cite{lapierre2020emergent} and \cite{EmergentSpacial} that the one cycle time translation is equivalent to a coordinate change given by a Möbius-transformation:
	\begin{align}
		\tilde{z}_1=\frac{az+b}{cz+d},
	\end{align}
	where ($\tau_0=\im T_0$ and $\tau_1=\im T_1$ are the imaginary periods):
	\begin{align}\label{eq: constants}
		\nonumber a&=\left(1+\frac{\pi \tau_1}{l}\right)\e^{\frac{\pi\tau_0}{l}} &b&=-\frac{\pi \tau_1}{l}\e^{\frac{-\pi\tau_0}{l}}\\c&=\frac{\pi \tau_1}{l}\e^{\frac{\pi\tau_0}{l}} &d&=\left(1-\frac{\pi \tau_1}{l}\right)\e^{\frac{-\pi\tau_0}{l}}
	\end{align}
	Since the fields are primary this results in:
	\footnotesize
	\begin{align}
		G(x,T,x_0,0)=\left(\frac{2\pi}{l}\right)^{h+\overline{h}}\Bigg(\frac{\partial \tilde{z}_1}{\partial z_1}\Bigg)^h\Bigg(\frac{\partial \tilde{\overline{z}}_1}{\partial \overline{z}_1}\Bigg)^{\overline{h}}\left<\phi(\tilde{z}_1,\tilde{\overline{z}}_1)\phi(z_0,\overline{z}_0)\right>
	\end{align}
	\normalsize
	By conformal symmetry, the last two-point function can be determined when particular boundary conditions are imposed \cite{PaulGinspargCFT}. For periodic boundaries we have:
	\begin{align}\label{eq: one cycle two point}
	\nonumber G(T,x,0,x_0)=&\left(\frac{2\pi}{l}\right)^{h+\overline{h}}\Bigg(\frac{\partial \tilde{z}_1}{\partial z_1}\Bigg)^h\Bigg(\frac{\partial \tilde{\overline{z}}_1}{\partial \overline{z}_1}\Bigg)^{\overline{h}}\\&\cdot \frac{1}{\big(\tilde{z}_1-z_0\big)^{2h}\left(\tilde{\overline{z}}_1-\overline{z}_0\right)^{2\overline{h}}}
	\end{align}
	It is possible to find the two-point function for stroboscopic times ($t=nT$ for some integer $n$) by applying the same transformation $n$ times. Denote the two fixed points of the Möbius-transformation as $\gamma_-$ and $\gamma_+$:
	\begin{align}\label{eq: gammas}
	\gamma_\pm=\frac{a-d\pm\sqrt{(a-d)^2+4bc}}{2c}
	\end{align}
	 A short proof of induction then gives that the coordinate transformation after $n$ cycles obeys:
	\begin{align}\label{eq: ztilden}
	\frac{\tilde{z}_{n}-\gamma_-}{\tilde{z}_{n}-\gamma_+}=\eta^{n}\frac{z-\gamma_-}{z-\gamma_+},
	\end{align}
	where:
	\begin{align}\label{eq: eta}
	\eta=\frac{a+d+\sqrt{(a-d)^2+4bc}}{a+d-\sqrt{(a-d)^2+4bc}}
	\end{align}
	Or by isolating $\tilde{z}_{n}$:
	\begin{align}\label{eq: n times}
	\tilde{z}_n=\frac{(\gamma_--\gamma_+\eta^n)z+\gamma_-\gamma_+(\eta^n-1)}{(1-\eta^n)z+\gamma_-\eta^n-\gamma_+}=\frac{\alpha z+\beta}{\delta z+\epsilon}
	\end{align}
	By using $\tilde{z}_n$ in equation \ref{eq: one cycle two point} instead of $\tilde{z}_1$ the stroboscopic two-point function has been found.
	
	\subsubsection*{Heating and Non-Heating Phases}
		\addcontentsline{toc}{section}{\protect Heating and Non-Heating Phases} 
	Even though the expression for the real-time two-point function is yet to be found, it is already possible to consider the overall behavior of the system, and whether or not it shows divergent or periodic tendencies. If $|\eta|\neq1$ then the system will converge and this phase will be called the heating phase. On the other hand, if $|\eta|=1$ but $\eta\neq 1$ then the system is periodic called the non-heating phase. Finally, if $\eta=1$ the equations cannot be applied since $\gamma_-=\gamma_+$ in this case. This is called the critical phase. The deciding factor of the form of $\eta$ is the sign of the number inside the square roots (multiplied by $\frac{1}{4}$ since only the sign matters):
	\begin{align}
	\Delta=\frac{1}{4}(a-d)^2+bc
	\end{align}
	Using $\tau\rightarrow \im t$, the result is calculated to be:
	\small
	\begin{align}
	\Delta=\left(\frac{\pi^2T_1^2}{l^2}-1\right)\sin^2\left(\frac{\pi T_0}{l}\right)-\frac{\pi T_1}{l}\sin\left(\frac{2\pi T_0}{l}\right)
	\end{align}
	\normalsize
	The expression is periodic in $T_0$, which is the driving period connected to the ordinary Hamiltonian $H_0$. For visualization, the leftmost plot in figure \ref{fig: DC} shows the two-point function only driven by $H_0$, and it becomes clear why periodicity occurs in this part of the driving.

\begin{figure}
\centering
\includegraphics[width=0.8\linewidth]{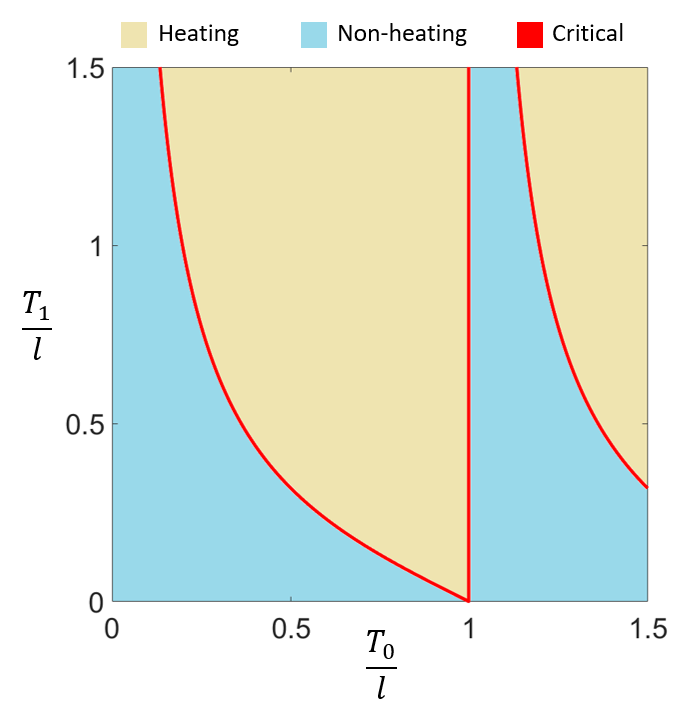}
\caption{The phase diagram of the driving protocol. The heating, non-heating, and critical driving phases can be seen. For more details on what happens within each phase, see \cite{lapierre2020emergent}. The driving is periodic in $T_0$ with period $l$. In fact, the entire system has oscillation period $l$ when driven by $H_0$ which explains the periodicity of the phase diagram. }
\label{fig: phase diagram}
\end{figure}

To see how $\Delta$ affects $\eta$, first realize that the $a+d$ in the expression for $\eta$ is a real number (also after the imaginary time continuation) since $a=\overline{d}$, where for once the bar actually denotes the complex conjugate. If $\Delta>0$, then the square root of $\Delta$ is also a real number, and $\eta\neq 1$. It is real, however. Consequently, $\Delta>0$ results in the heating-phase. If $\Delta<0$, the square roots become imaginary and $\eta$ is a complex number. However, the norm of $\eta$ is one thus resulting in the non-heating phase. If $\Delta=0$, the driving is critical, and the system behaves like some kind of a mix between heating and non-heating. For example, the energy density of the system oscillates in the non-heating phase, grows exponentially around the points of convergence in the heating phase, and in the critical phase, the energy density still grows around the fixed points albeit much slower\footnote{At least, this is the case for stroboscopic times. Whether the same patterns emerge for real times is an open question.} \cite{EmergentSpacial}. Figure \ref{fig: phase diagram} shows the phase diagram depending on the two driving periods $T_0$ and $T_1$. It should be mentioned that the two-point function is highly sensitive to any changes in the driving periods when being close to the critical driving. This high sensitivity could either be exploited experimentally or should be avoided, depending on the scenario.

An advantage of working with the stroboscopic picture (omitting the micro-motion to be introduced in this work), is that it allows an effective Hamiltonian to be constructed which shows the different phases from its classification in terms of the Casimir of $SL(2,\mathbb{R})$ \cite{lapierre2020emergent}.

	\subsubsection*{The Real-Time Two-Point Function} 
	\addcontentsline{toc}{section}{\protect The Real-Time Two-Point Function}
	Since much information is to be found in the stroboscopic solution, the deduction of the real-time solution has not been carried out in the literature before. In fact, looking back at equation \ref{eq: ztilden}, it is easy to guess that the final result could be:
	\begin{align}
	\frac{\tilde{z}_t-\gamma_-}{\tilde{z}_t-\gamma_+}=\eta^{t/T}\frac{z-\gamma_-}{z-\gamma_+},
	\end{align}
	This assumption will be called method 1 and is usually implied in the referenced articles. There is a way to test if this method is correct, however. Consider an arbitrary real time $t$ and write it as:
	\begin{align}
		t=nT+r,
	\end{align}
	where $0\le r<T$ is the rest term and $n$ is a non-negative integer. If $r<T_0$ then this scenario is equivalent to driving the system up to $nT$ followed by driving the system during the time $r$ using the ordinary Hamiltonian. This is equivalent to transforming $z$ to $\tilde{z}_n$ by equation \ref{eq: n times} and then transforming $\tilde{z}_n$ again according to the same equation but using $T_0=r$ and $T_1=0$ to calculate the four constants of the transformation. If $r\ge T_0$ then the same method is applied, but the effective values of the driving periods in the last part are now $T_1=r-T_0$ and $T_0$ is untouched. 
	
	In other words, the transformation from $z$ to $\tilde{z}_t$ can be written as:
	\begin{align}\label{eq: zt}
		\tilde{z}_t=\frac{A\tilde{z}_n+B}{C\tilde{z}_n+D},
	\end{align}
	where $\tilde{z}_n$ can be found using equation \ref{eq: n times} and $A$, $B$, $C$, and $D$ are given by equations \ref{eq: constants}, but using the mentioned replacements for $T_0$ and $T_1$. This will be called method 2. 
	
\begin{figure}
\centering
\includegraphics[width=\linewidth]{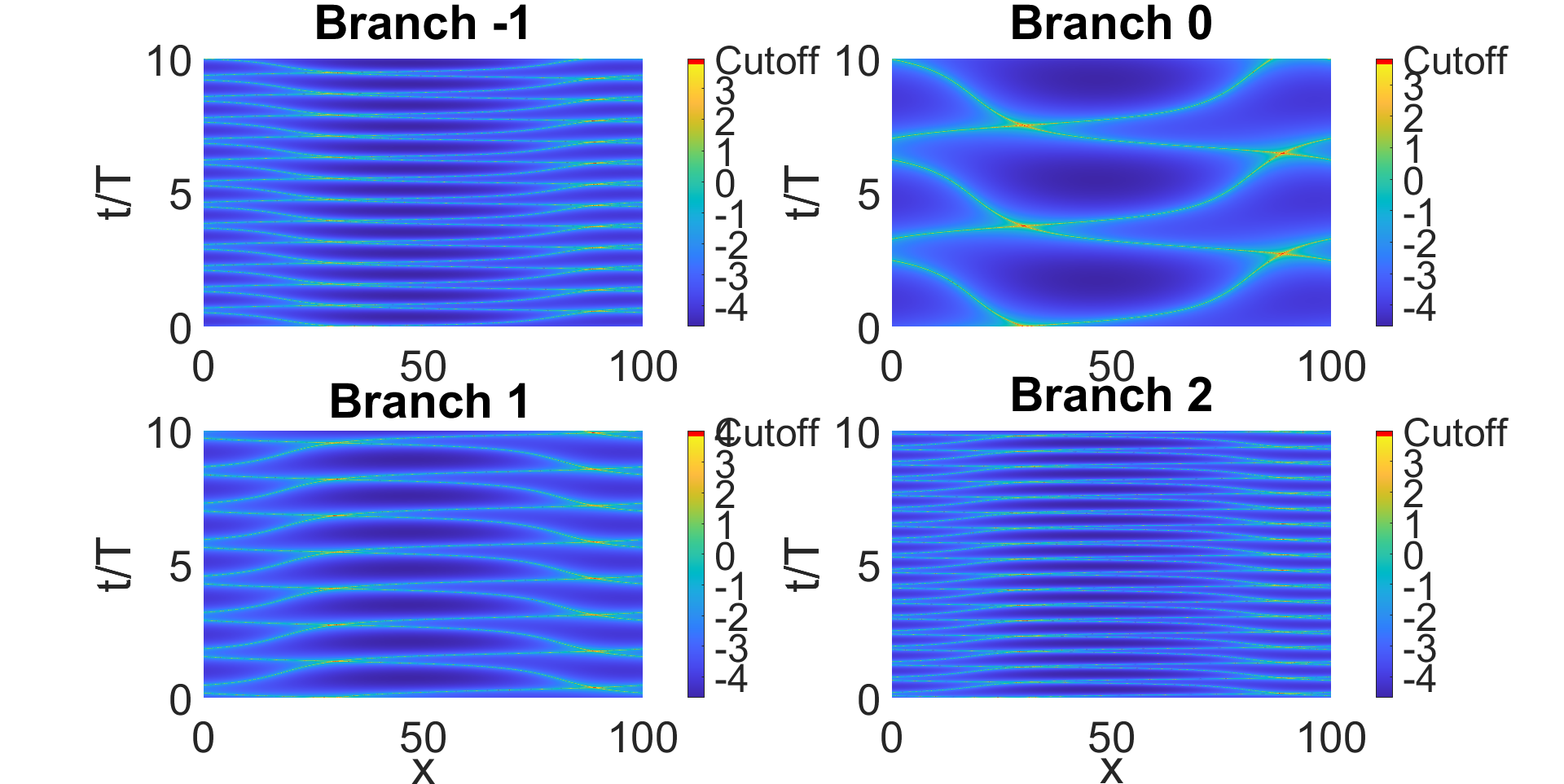}
\caption{The two-point function for different branches. The conditions are $T_0=37,T_1=37,l=100,x_0=30$ which results in $|\eta|=1$ and thus the non-heating phase. The further the branch is away from 0, the smaller the period of oscillation appears to be.}
\label{fig: branches}
\end{figure}

	Looking back at the equation for method 1 a question arises. Since $\eta$ is not necessarily real and the power $t/T$ is in general not integer anymore the result depends on the chosen branch. The $n$'th branch of raising a complex number $q$ to a power $p$ is defined as:
	\begin{align}
	q^p=\e^{p(\log(q)+\im 2n\pi)},\qquad n\in\mathbb{N},
	\end{align}
	where $\log$ is the principal logarithm. Of course, all branches in method 1, as well as method 2, agree for all stroboscopic times ($t/T\in\mathbb{Z}$). Figure \ref{fig: branches} shows a plot of $\log_{10}|G|$ for different branches. In all figures of this text, the condition $h=\overline{h}=1$ is used since these values have little impact on the overall pattern. A cutoff is also introduced since the two-point function diverges along certain lines. It can be seen that the result is highly dependent on the chosen branch which is problematic.

	\subsubsection*{Discussion of Method 1 and Method 2}
\begin{figure}
\centering
\includegraphics[width=\linewidth]{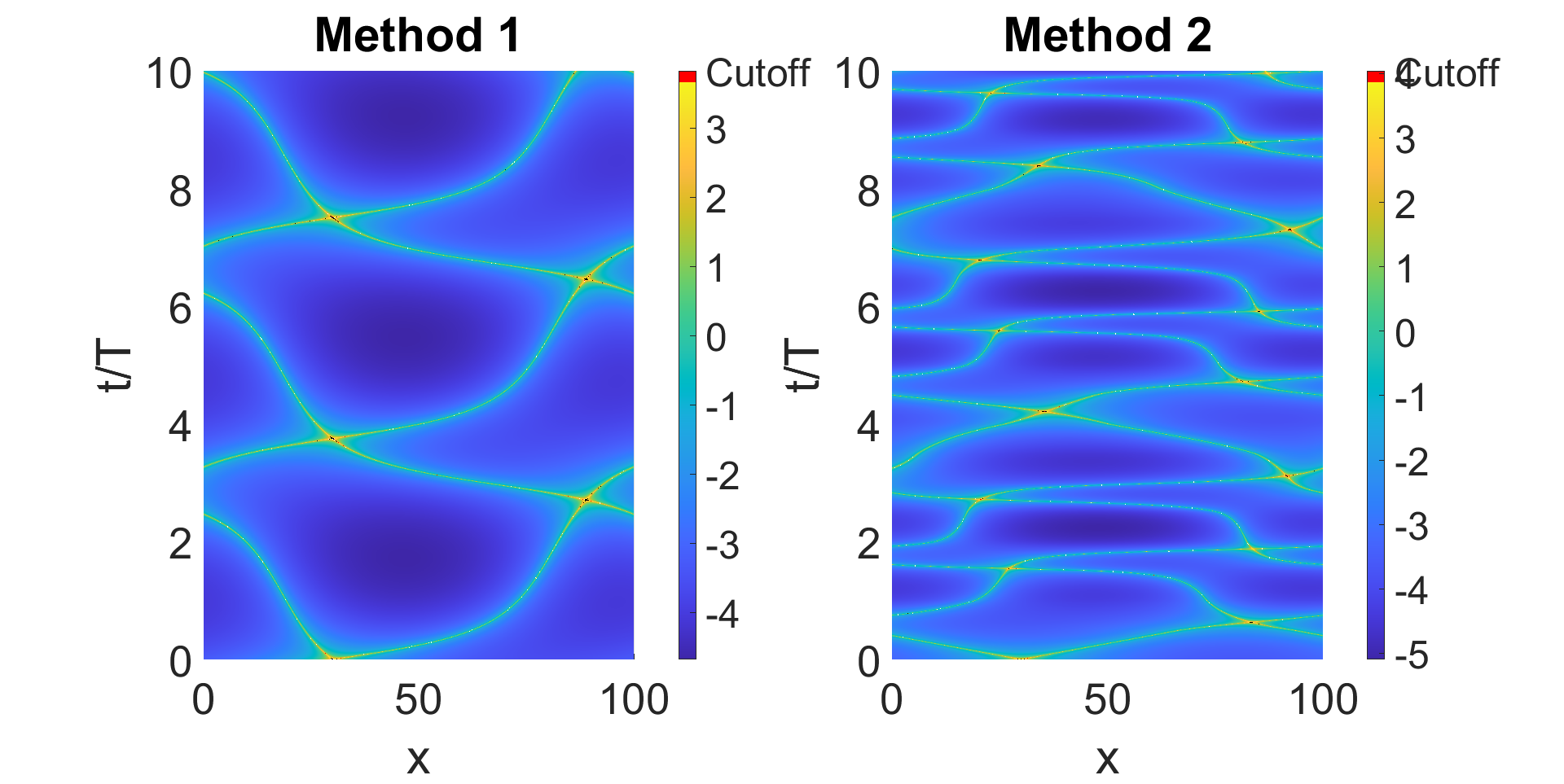}
\caption{The two-point function for method 1 (branch 0) and method 2 in the non-heating phase. The conditions are as earlier $T_0=37,T_1=37,l=100,x_0=30$ which results in $|\eta|=1$.}
\label{fig: comp1}
\end{figure}

		\addcontentsline{toc}{section}{\protect Discussion of Method 1 and Method 2}
Comparisons between methods are all shown for the same value of $x_0$ since this value has no particular effect when comparing the two results. The consequence of different values of $x_0$ is discussed briefly in the end of this section. Figure \ref{fig: comp1} shows a comparison between the two methods in the non-heating phase. It might not be clear from the figure that the two methods agree on stroboscopic times even though this is, of course, the case. Instead, the figure illustrates that the two methods disagree in between real times where method 2 takes the micro-motion of the quasi-particles into account. It should be mentioned that any other choice of branch is even worse in comparison as appears from figure \ref{fig: branches}, which has the same conditions. That branch 0 is the best choice will reappear in several later figures, but the reason is unknown.
		
There are similarities between the two methods though. Both methods have two trajectories of the quasi-particles. These are the two lines of maxima\footnote{These lines are where the function diverges as mentioned earlier. Divergence only occurs exactly on the lines, which means the finite number of points actually plotted are finite.} that appear in both cases such that one quasi-particle is moving to the left while another is moving to the right. Notice that the quasi-particles move through the boundaries, which is possible because we chose periodic boundary conditions for the conformal two-point function. These two effects are generic in periodic Floquet CFT with the sine-square deformation\footnote{One of the effects of SSD is that the ground states of the original and deformed Hamiltonian agree, \textit{if} periodic boundary conditions of the \textit{original} (not the deformed) system are assumed\cite{katsura}.}.

\begin{figure*}
\centering
\includegraphics[width=\linewidth]{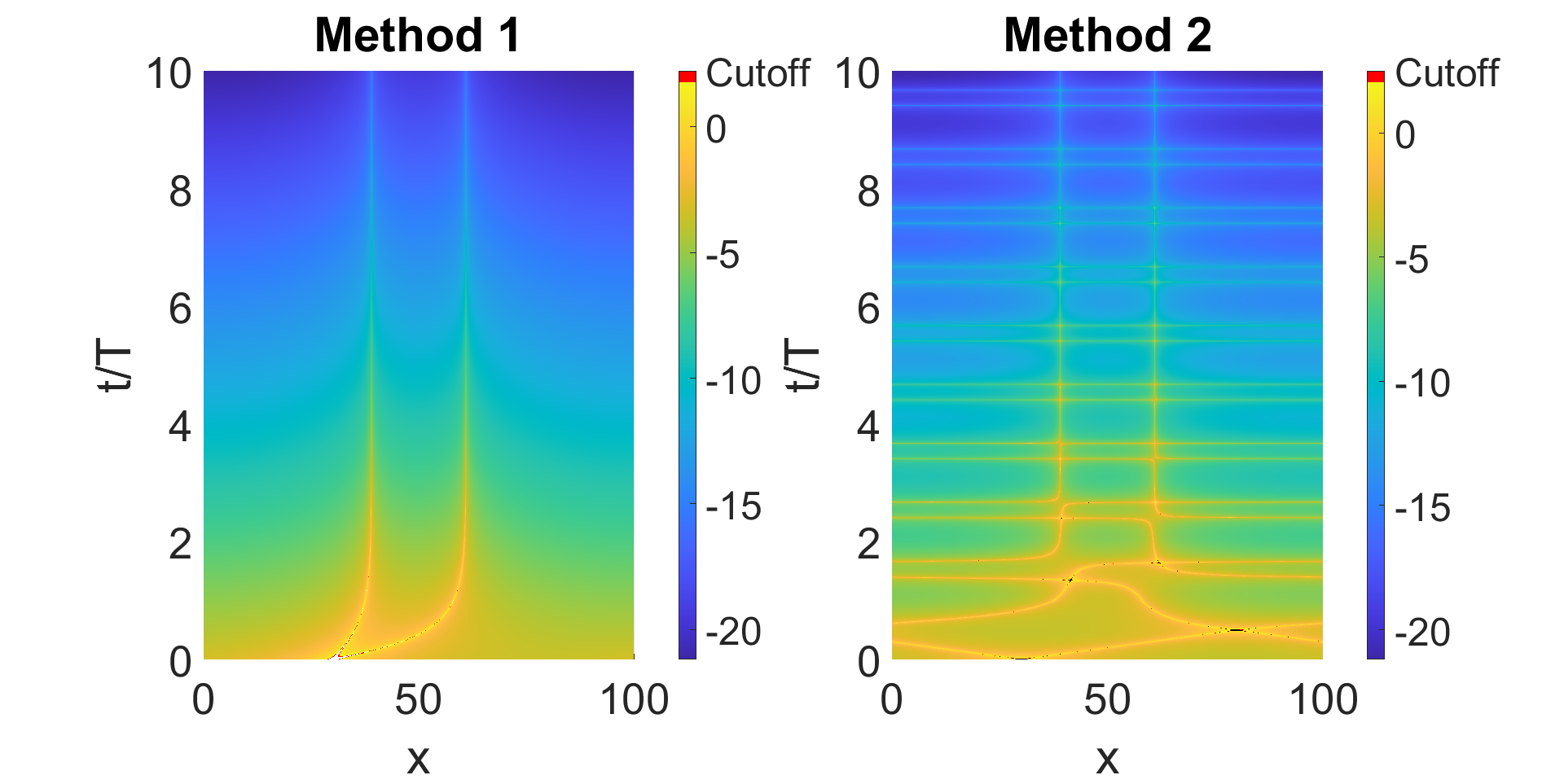}
\caption{The two-point function for method 1 (branch 0) and method 2 in the heating phase. The conditions are $T_0=50,T_1=50,l=100,x_0=30$ which results in $|\eta|\approx 0.13$.}
\label{fig: comp4}
\end{figure*}

A more interesting comparison occurs in the heating phase illustrated in figure \ref{fig: comp4}. Ignoring the ''horizontal'' lines\footnote{These almost horizontal lines are a consequence of the $H_0$-driving. We will later see that method 1 seems to catch the effects of $H_0$ poorly while the effects of $H_{\text{SSD}}$ are captured quite well.} of method 2, the two methods actually agree quite well. In \cite{lapierre2020emergent}, the trajectories of the quasi-particles as described by method 1 (stroboscopic extrapolation) was shown to follow light-like geodesics in a 2-dimensional Schwarzschild-metric. The light-like geodesic is expected if any since conformal field theories are massless. That is, the quasi-particle trajectories converge towards two points which act approximately like black holes. This will be discussed later. However, when applying method 2 (including micro-motion) convergence does not happen. Instead, the quasi-particles \textit{slows down} when approaching the points of convergence and speed up when leaving them\footnote{The trajectory going to the left is only affected by the right point of stroboscopic convergence and similarly for the other line.}. This means that the physical interpretation of the convergence points has turned from attraction to repulsion. It also means that the black hole interpretation is not adequate when including the real time micro-motion of the quasi-particles, even though the relation to black holes still exists at all stroboscopic times. 

The ''strength'' and ''range'' of the repulsion is reflected in how fast the quasi-particle trajectories slow down and accelerate around the points. It can be seen that the ''strength'' seems to increase with time while the ''range'' is decreasing since the deceleration and acceleration become rapidly more intense with time and happens in a smaller region (the spatial region could also be smaller because the line is moving faster). Remember, that the slope of a trajectory is the inverse of the quasi-particles speed. Thus, if the slope changes mostly within a small time interval, the quasi-particle acceleration must be large, and if the slope changes mostly within a small spatial interval, the range must be short. Note, that these are qualitative observations only and that the idea of strength and range is at most a convenient picture to keep in mind.

\begin{figure}
\centering
\includegraphics[width=\linewidth]{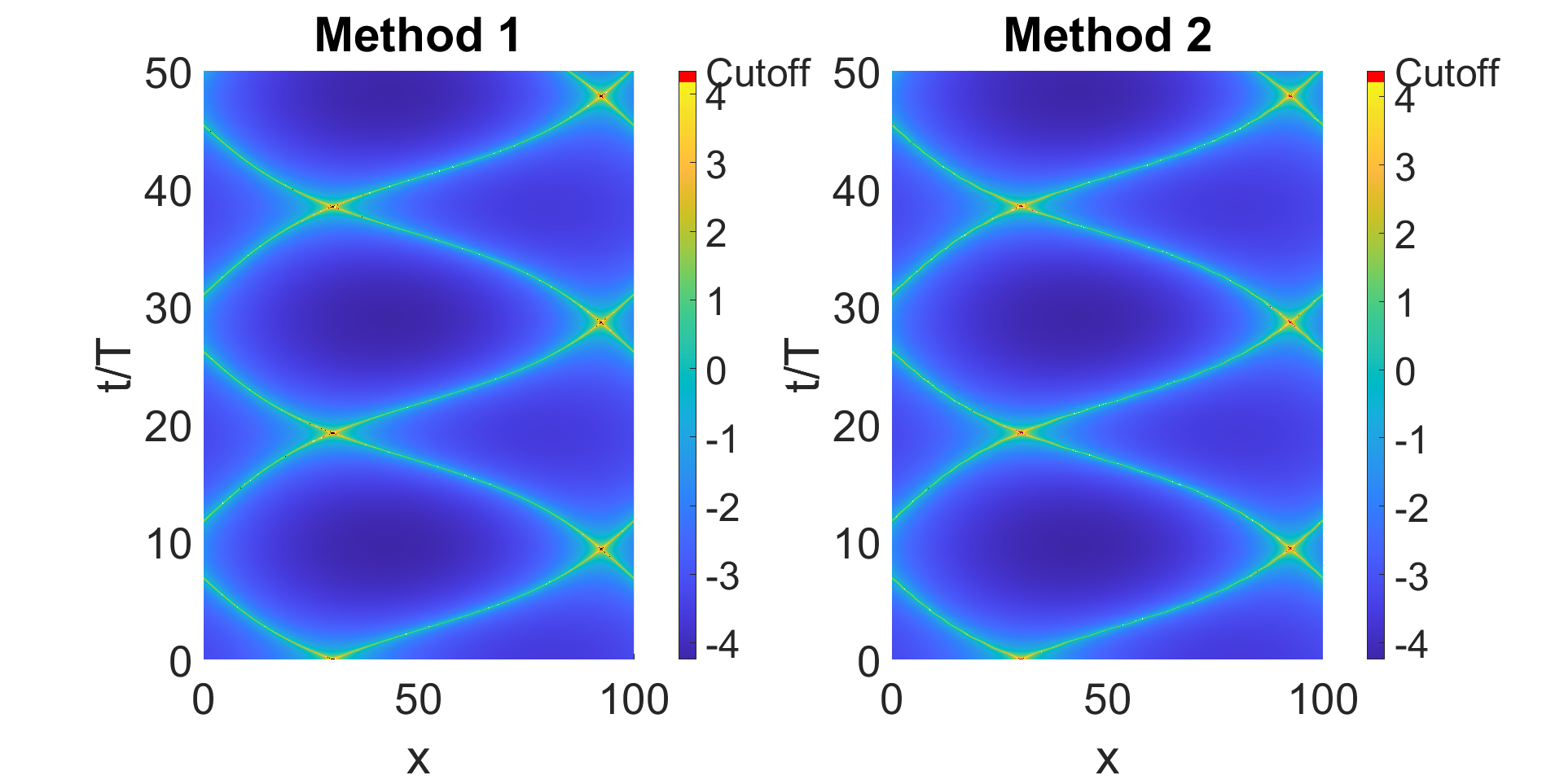}
\caption{The two-point function for method 1 (branch 0) and method 2 in the non-heating phase. The conditions are $T_0=3,T_1=3,l=100,x_0=30$ which results in $|\eta|=1$. The function is plotted over more periods than earlier.}
\label{fig: comp3}
\end{figure}

It is important to investigate if method 2 really is the correct method. Numerical simulations have already been made in other articles confirming the predictions of method 1 (one such simulation of a spin chain was made in \cite{lapierre2020emergent}). Interestingly, there are regimes where the two methods agree to high accuracy, and the simulations have been carried out in this regime in \cite{lapierre2020emergent}. An example can be seen in figure \ref{fig: comp3}. It is worth finding the regimes in which method 1 is a good approximation for three reasons. Firstly, the underlying and yet unknown explanation might help discover new phenomena or describe the system better.

 Secondly, method 1 is simpler analytically, especially since method 2 has to be split into two different cases to carry out the calculations. Thirdly, if any future simulations are to be carried out distinguishing between the two methods, this simulation must be carried out in the right regime.

\begin{figure}
\centering
\includegraphics[width=\linewidth]{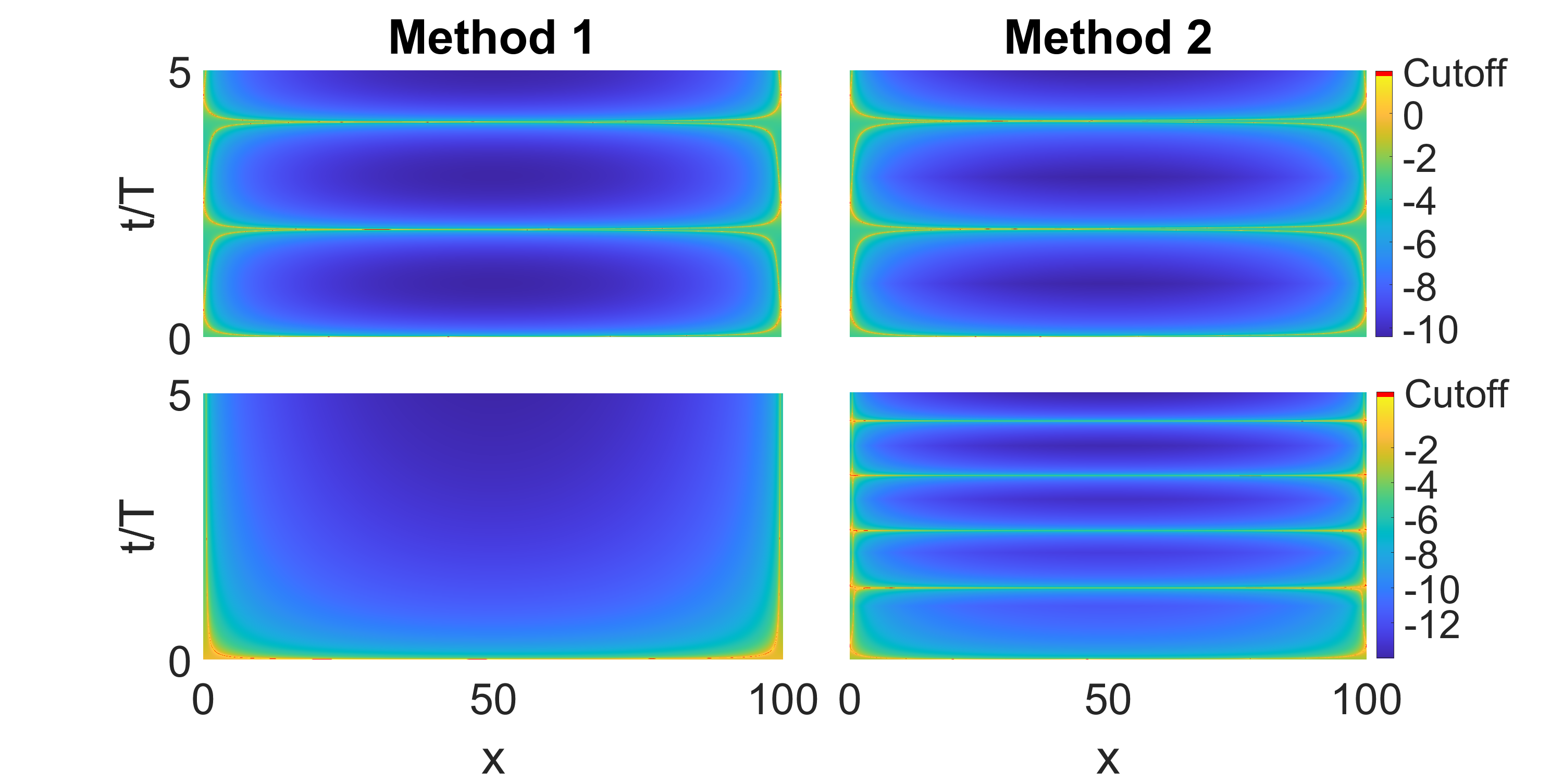}
\caption{$T_0=1$ and $x_0=30$ on all four plots. They are all in the non-heating phase, however, the bottom left and bottom right are close to the critical driving at $T_1\approx 2.03\cdot 10^3$. Top left and top right: $T_1=10^3$. Bottom left and bottom right: $T_1=2\cdot 10^3$.}
\label{fig: comp5}
\end{figure}

 It appears that two criteria need to be satisfied for the methods to approximately agree. Firstly, the period $T_0$ has to be small compared to $l$, meaning that the $H_0$-driving happens in short intervals\footnote{Though it can still make up most of the total driving time. In figure \ref{fig: comp3} the driving of $H_0$ lasts 50\% of the entire driving. }. For instance, this is the case in figure \ref{fig: comp3} since $T_0/l=0.03$. It is clear why this is the case. When $T_0$ and $T_1$ are small (the high frequency regime), there is little difference between the stroboscopic dynamics and the real-time dynamics since the time difference, $T=T_0+T_1$, between stroboscopic points becomes small. When $T$ becomes small enough, so that is comparable with the resolution of the real-time parameter, the two methods produce identical results. It is clear from the phase diagram in figure \ref{fig: phase diagram} that the high-frequency regime is non-heating. However, heating phase dynamics can be obtained in this regime by changing the sign of the uniform Hamiltonian $H_0$. It also seems to be important that the driving is ''well within'' the non-heating phase, which means that $T_1$ cannot be too large so it comes close to the critical line in figure \ref{fig: phase diagram}, even though it can still be large. This indicates that method 1 is good at describing time evolution by $H_\text{SSD}$, but not the evolution governed by $H_0$. If this is true, the same argument as before applies here as well since no considerable differences between the two methods have time to develop during the short driving by $H_0$, and no or at least very small changes occur during the $H_\text{SSD}$-driving. However, when the system approaches or even enters the heating regime, very rapid changes take place over time. Now, remember the term ''small'' is compared to the time-scale of the system, which means that ''$T_0$ has to be small'' is a very restrictive constraint in these circumstances. This would explain why $T_0=1$ in the bottom of figure \ref{fig: comp5} can no longer be regarded as small enough for method 1 to be a good approximation.

 Figure \ref{fig: comp5} shows some extreme cases and illustrates that method 1 is a good approximation even for very large $T_1$ as long as $T_0$ is small, the driving is in the non-heating phase, and far away from the critical driving. Method 1 is never a good approximation in the heating phase since the ''horizontal'' lines appear with method 2 but never with method 1. Interestingly, in the heating phase, $\eta$ is real and the entire problem about branches does therefore not exist here. All the terms ''well within'' and ''far away'' cannot be stated more precisely since the exact reasons for the similarities are not yet understood.
 
 The fact that $T_0$ has to be small indicates that method 1 is bad at modelling $H_0$-driving. For $t<T_0$, the expression for the two-point function is simple since for such times no actual driving has been introduced. The exact analytical result for $t<T_0$ is
 \begin{align}\label{eq: two point H0}
 &G(t,x,0,x_0)=\left(\frac{2\pi}{l}\right)^{h+\overline{h}}\e^{\frac{\im 2\pi (h+\overline{h})}{l}t}\\
 \nonumber &\cdot\frac{1}{\left(\e^{\frac{\im 2\pi}{l}(t+x)}-\e^{\frac{\im 2\pi}{l}x_0}\right)^{2h}}
 \frac{1}{\left(\e^{\frac{\im 2\pi}{l}(t-x)}-\e^{-\frac{\im 2\pi}{l}x_0}\right)^{2\overline{h}}}
 \end{align}
This expression gives the $t<T_0$ quasi-particle trajectories:
\begin{align}\label{eq: lines}
x+t=x_0\mathand x-t=x_0
\end{align}
This explains why the period, when the system is described by $H_0$, is exactly $l$. The expression is of course more complicated after more periods of driving, but this fact seems to be conserved even though the trajectories are no longer straight lines. 

\begin{figure}
	\centering
	\includegraphics[width=\linewidth]{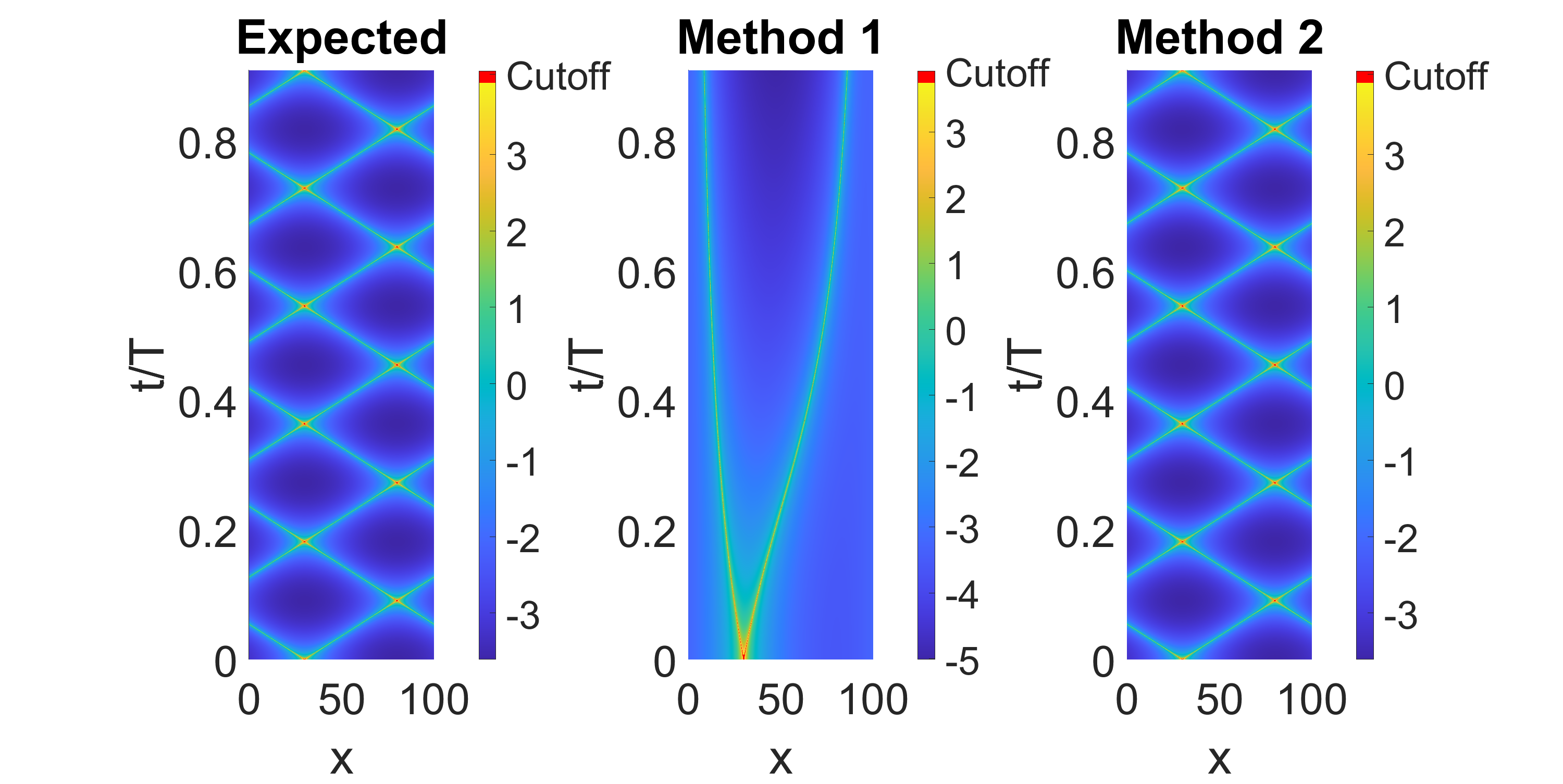}
	\caption{The final comparison between the two methods. In this case, the periods are $T_0=500,T_1=50$ and $x_0=30$, however, only the first part of the driving is shown $(0\le t\le T_0)$. }
	\label{fig: DC}
\end{figure}

The result in this time-span is plotted in figure \ref{fig: DC} and illustrates that method 1 completely fails to describe driving by $H_0$ while method 2 yields the correct result.

\begin{figure}
	\centering
	\includegraphics[width=\linewidth]{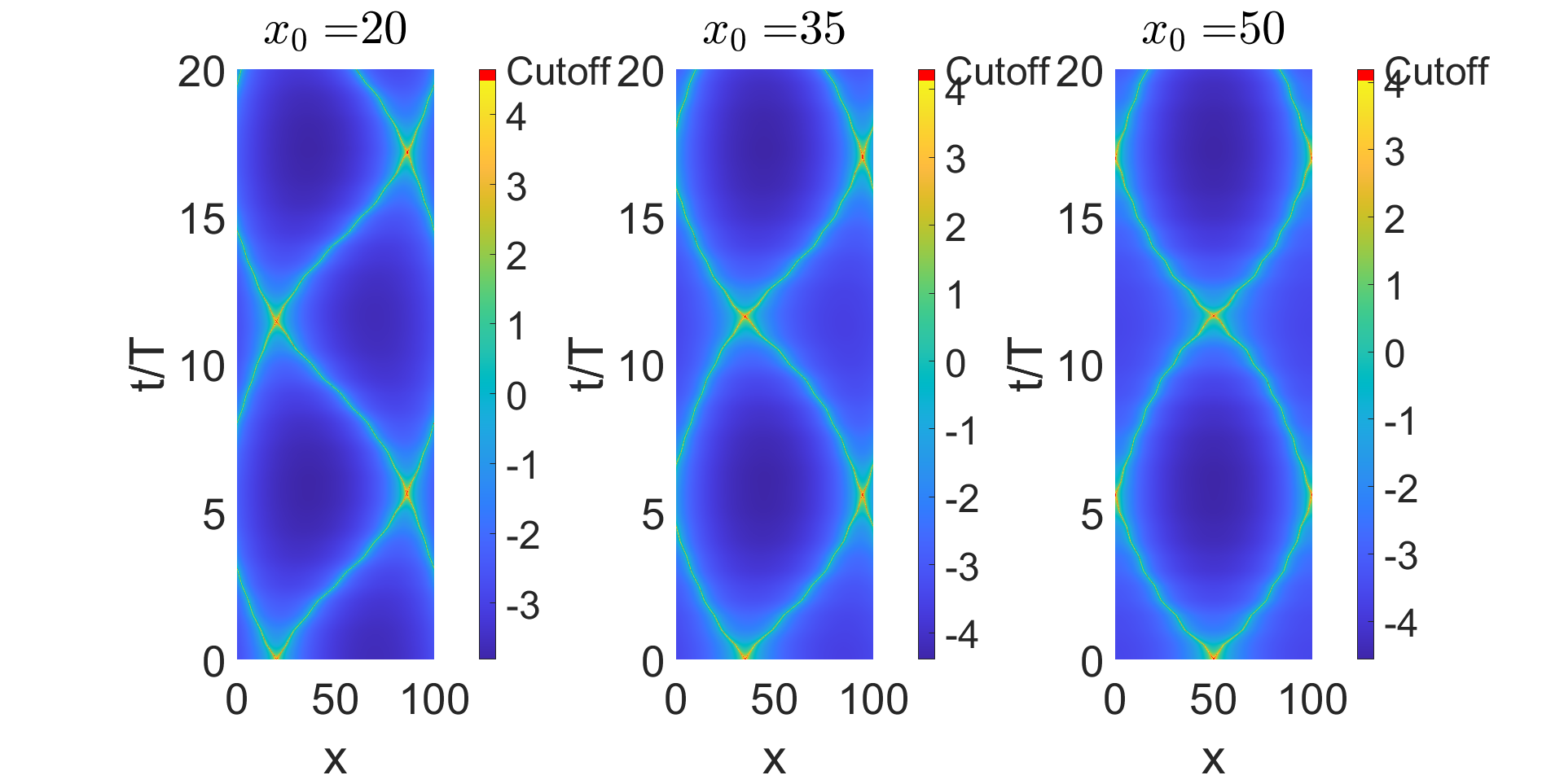}
	\caption{Plots for method 2 in the non-heating phase with $T_0=5$ and $T_1=5$. The situation for $x_0>l/2$ is symmetric to $x_0<l/2$.}
	\label{fig: InCondNon}
\end{figure}

\begin{figure}
	\centering
	\includegraphics[width=\linewidth]{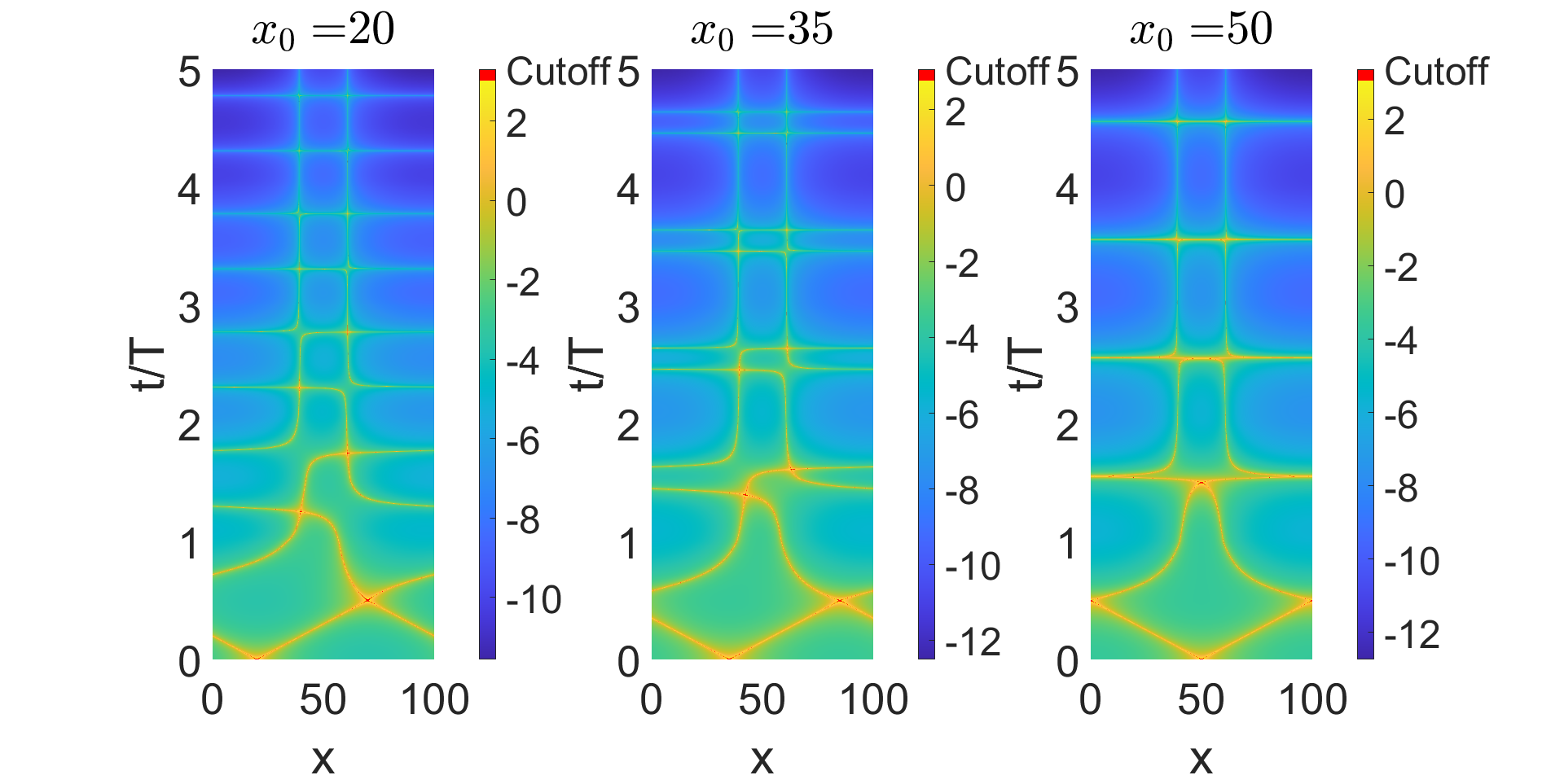}
	\caption{Plots for method 2 in the heating phase with $T_0=50$ and $T_1=50$. The situation for $x_0>l/2$ is symmetric to $x_0<l/2$.}
	\label{fig: InCondHeat}
\end{figure}

Lastly, the two figures \ref{fig: InCondNon} and \ref{fig: InCondHeat} illustrates the effect of different initial conditions. That is different values of $x_0$. The situation is symmetric for $x_0>l/2$. In the non-heating phase, the patterns are the same but simply ''weighted'' towards the position $x_0$. In the heating phase, the vertical lines are at the same position independent on $x_0$ since they have to agree with method 1 in the stroboscopic times. However, the horizontal lines show a greater vertical distance between the two quasi-particles when the distance to the center $|x_0-l/2|$ increases. 

\subsubsection*{The Geometric Interpretation}
\addcontentsline{toc}{section}{\protect The Geometric Interpretation}

\begin{figure}
\centering
\includegraphics[width=\linewidth]{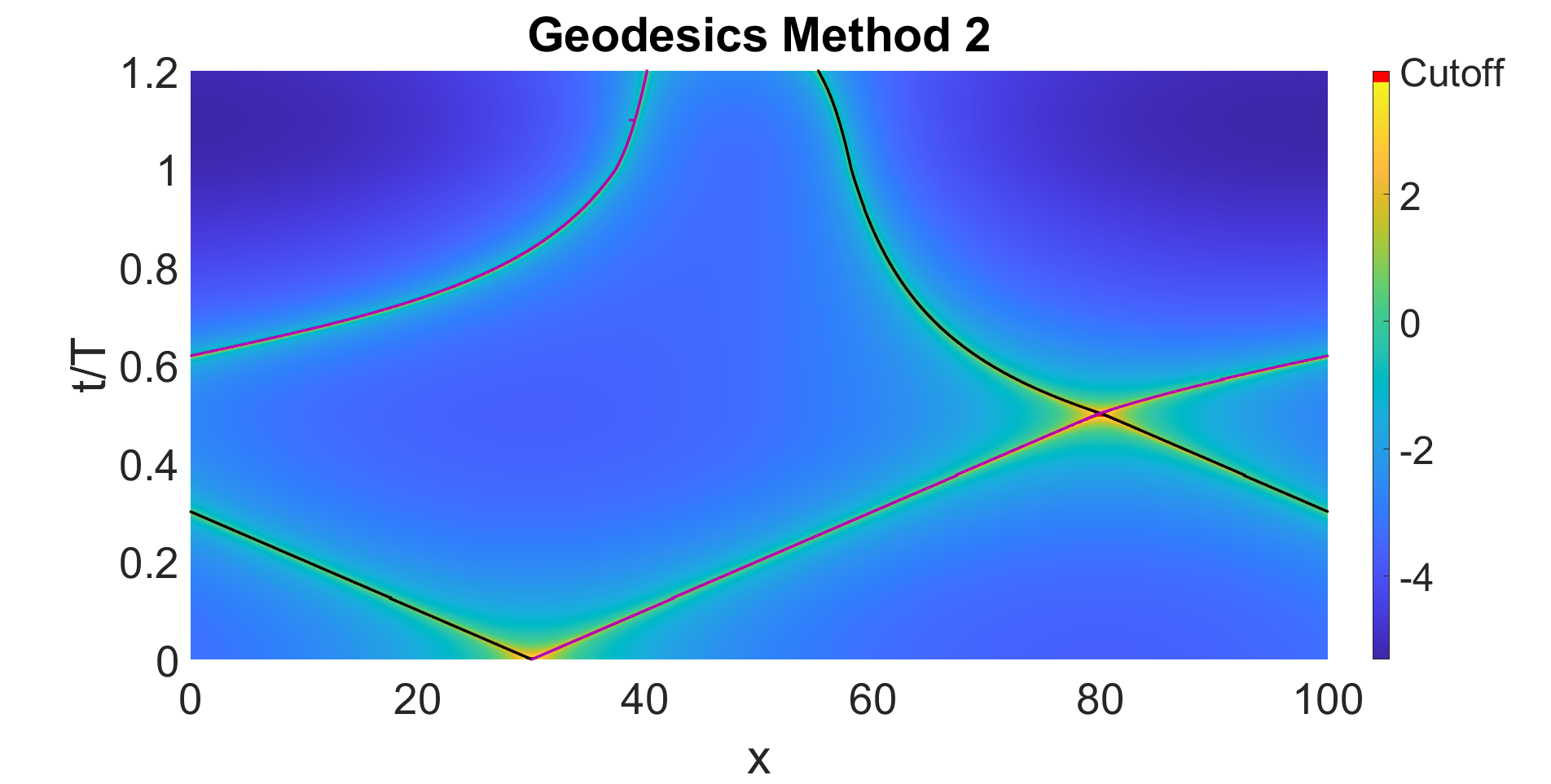}
\caption{A numerical simulation of solutions to $\diff s^2=\diff \tilde{z}_t\diff \tilde{\overline{z}}_t=0$ in the heating-phase. There are two solutions. The two solutions are plotted in purple and black on top of the two-point function driven by $T_0=T_1=50$ (the same conditions as in figure \ref{fig: comp4}).}
\label{fig: NumSimGeo}
\end{figure}

In \cite{lapierre2020emergent}, the surprising result that there exists a relation to black hole dynamics in SSD-driven systems in the heating-phase was discovered. However, since the real time motion was found by method 1, this result has to be questioned. The article \cite{lapierre2020emergent} argues that the coordinates $(\tilde{z}_n,\tilde{\overline{z}}_n)$ given by equation \ref{eq: n times} are \textit{isothermal} coordinates of an effective manifold on which the light-like geodesics agree with the lines of maxima. That is, the metric is given by:
\begin{align}
\diff s^2=\diff \tilde{z}_n\diff \tilde{\overline{z}}_n,
\end{align}
and the lines of maxima solve the light-like\footnote{Recall, that conformal fields are massless and thus must follow the light-like geodesic if any.} equation:
\begin{align}
\diff s^2=0
\end{align}
This idea can be generalized to agree with method 2 simply by taking the metric as:
\begin{align}
\diff s^2=\diff \tilde{z}_t\diff \tilde{\overline{z}}_t,
\end{align}
where $\tilde{z}_t$ is given by equation \ref{eq: zt}. The equations of motion, and thus the light-like curves, are of course different. In fact, in this case, they correspond to the result obtained by method 2 showing that the geometrical argument used in \cite{lapierre2020emergent} still works. A numerical simulation is shown in figure \ref{fig: NumSimGeo} and compared to the two-point function in the heating-phase. The equations are way more complicated, however, and are therefore no longer dealt with analytically, contrary to the equations obtained when using method 1. 

This is another example of why method 1 still has benefits, although it does not describe the real-time correctly. It also demonstrates how some ideas applied to method 1 still works when applied to method 2.

Of course, the interpretation of real-time motion is no longer that of black holes. But the real-time motion can be stroboscopically interpreted as black hole motion close to the critical points. It is interesting that this interpretation of black holes somehow emerges in the very complex real-time motion, but only at the end of each driving cycle. 

It should be mentioned that the current differential equations are very challenging to work with, even numerically. Since the solutions give the trajectories of the quasi-particles an easier way to find them is to find the two-point function (which has the extremely rare benefit of having a known analytical solution) and then simply tracking the trajectories. This also has the consequence that no numerical uncertainty will add up for long time simulations and that the duration of the simulation is unaffected by the desired time span.

\subsubsection*{Conclusions}
	\addcontentsline{toc}{section}{\protect Conclusion}
	
Sine-square deformed driven conformal field theories have received much attention recently, but the results have only been calculated up to stroboscopic times, while the full real-time results have merely been implied as an extrapolation thereof. We have found that extrapolation to real times must be done with great care to obtain reliable results. We present an exact method that applies for all times that solves the issue with extrapolation. Though the original extrapolation to real times is not generally reliable, there exists regimes in which it is a good approximation. We have shown that the limit where $T_0\ll l$ and $T_1$ is small enough to stay well within the non-heating phase, shows the same behavior in both cases. We argue that this result originates from the fact that the old extrapolation from stroboscopic times accounts poorly for the dynamics of $H_0$.

Furthermore, as we have discussed, the proposed relationship between SSD conformal Floquet dynamics and the geodesic of a particular black hole dynamics \cite{lapierre2020emergent} holds only for stroboscopic times, though the geometrical interpretation still works for the new method presented here. The new geodesic equations are too complicated and we have not found analytical solutions, even in appropriate limits. We have made a numerical simulation of the new geodetic motion and compared it to the new two-point function, and our results show excellent agreement between the two approaches. We thus conclude that while the geometric indeed displays the hallmark of a curved space-time, it is not a simple matter to characterize the features of this spacetime. This is an interesting question for future research.

\begin{acknowledgments}
We would like to thank Bastien Lapierre and Ramasubramanian Chitra for looking through our manuscript and for giving us their useful comments.
\end{acknowledgments}

\bibliography{bib-method1.bib}

\end{document}